\documentclass[sigconf]{acmart}
\usepackage{www24-llm-adblocker-frame}
\graphicspath{{../www24-llm-adblocker-figures/}{./}}

\begin{document}
\title{Detecting Generated Native Ads in Conversational Search}

\author{Sebastian Schmidt}
\orcid{0000-0002-7701-3294}
\affiliation{%
\institution{Leipzig University}
\city{}
\country{}
}

\author{Ines Zelch}
\orcid{0009-0005-2659-5326}
\affiliation{%
\institution{Leipzig University and Friedrich-Schiller-Universität Jena}
\city{}
\country{}
}

\author{Janek Bevendorff}
\orcid{0000-0002-3797-0559}
\affiliation{%
\institution{Leipzig University and Bauhaus-Universität Weimar}
\city{}
\country{}
}

\author{Benno Stein}
\orcid{0000-0001-9033-2217}
\affiliation{%
\institution{Bauhaus-Universität Weimar}
\city{}
\country{}
}

\author{Matthias Hagen}
\orcid{0000-0002-9733-2890}
\affiliation{%
\institution{Friedrich-Schiller-Universität Jena}
\city{}
\country{}
}

\author{Martin Potthast}
\orcid{0000-0003-2451-0665}
\affiliation{%
\institution{Leipzig University and ScaDS.AI}
\city{}
\country{}
}

\renewcommand{\shortauthors}{Sebastian Schmidt et al.}

\begin{abstract}
Conversational search engines such as YouChat and Microsoft Copilot use large language models (LLMs) to generate responses to queries. It is only a small step to also let the same technology insert ads within the generated responses---instead of separately placing ads next to a response. Inserted ads would be reminiscent of native advertising and product placement, both of which are very effective forms of subtle and manipulative advertising. Considering the high computational costs associated with LLMs, for which providers need to develop sustainable business models, users of conversational search engines may very well be confronted with generated native ads in the near future. In this paper, we thus take a first step to investigate whether LLMs can also be used as a countermeasure, i.e., to block generated native ads. We compile the Webis Generated Native Ads~2024 dataset of queries and generated responses with automatically inserted ads, and evaluate whether LLMs or fine-tuned sentence transformers can detect the ads. In our experiments, the investigated LLMs struggle with the task but sentence transformers achieve precision and recall values above~0.9.
\end{abstract}



\copyrightyear{2024}
\acmYear{2024}
\setcopyright{rightsretained}
\acmConference[WWW '24 Companion]{Companion Proceedings of the ACM Web Conference 2024}{May 13--17, 2024}{Singapore, Singapore}
\acmBooktitle{Companion Proceedings of the ACM Web Conference 2024 (WWW '24 Companion), May 13--17, 2024, Singapore, Singapore}\acmDOI{10.1145/3589335.3651489}
\acmISBN{979-8-4007-0172-6/24/05}

\makeatletter
\gdef\@copyrightpermission{
   \begin{minipage}{0.3\columnwidth}
     \href{https://creativecommons.org/licenses/by-sa/4.0/}{\includegraphics[width=0.90\textwidth]{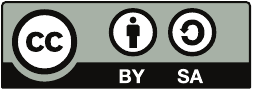}}
   \end{minipage}\hfill
   \begin{minipage}{0.7\columnwidth}
     \href{https://creativecommons.org/licenses/by-sa/4.0/}{This work is licensed under a Creative Commons Attribution-ShareAlike International 4.0 License.}
   \end{minipage}
   \vspace{5pt}
}
\makeatother

\maketitle

\section{Introduction}

Large language models~(LLMs) have quickly become the de facto standard for building conversational search engines and retrieval-augmented generation systems. Still, deploying LLM-based search engines at scale is expensive, while it is not yet clear what the best business model is for their sustainable operation. Subscription models are conceivable but given that advertising is very profitable and widely used in traditional search engines~\cite{jafarzadeh:2015,lewandowski:2018}, it is likely that ads will also play a role in conversational search. Some early announcements from Microsoft%
\footnote{\raggedright \href{https://web.archive.org/web/20230330001739/https://blogs.bing.com/search/march_2023/Driving-more-traffic-and-value-to-publishers-from-the-new-Bing}{blogs.bing.com/search/march\_2023/Driving-more-traffic-and-value-to-publishers}}
and Google%
\footnote{\raggedright \href{https://web.archive.org/web/20230523173443/https://blog.google/products/ads-commerce/ai-powered-ads-google-marketing-live}{blog.google/products/ads-commerce/ai-powered-ads-google-marketing-live}}
provided first insights into their ongoing developments in this regard.

Conversational search engines open up new opportunities for advertising, as they can insert ads for products or brands relevant to a query directly into their generated responses. Similar forms of marketing are already known as native advertising, where sponsored content is designed to resemble the style of non-commercial content~\cite{amazeen:2020,wojdynski:2016}, and product placement, where products are seamlessly inserted and shown, for instance, as part of a piece of entertaining content. Various trade and media regulations, e.g., from the United States Federal Trade Commission, require appropriate disclosure of such ads to the consumers.%
\footnote{\raggedright\href{https://web.archive.org/web/20230530021453/https://www.ftc.gov/system/files/documents/public_statements/896923/151222deceptiveenforcement.pdf}{\mbox{ftc.gov/system/files/documents/public\_statements/151222deceptiveenforcement.pdf}}\hspace*{-1pt}}

Still, under current ad disclosure standards, the majority of users do not seem to be able to recognize native advertising~\cite{amazeen:2020} nor to reliably distinguish between paid content and organic search results~\cite{lewandowski:2017}. One reason is that the proverbial line between ads and organic web search results is often blurry on traditional results pages, presumably with the intention of maximizing the number of clicks on paid results~\cite{lewandowski:2017,lewandowski:2018}. Inserting an ad natively into a generated response could further increase the difficulty of recognizing paid content, making users more susceptible to manipulation~\cite{amazeen:2020}.

In this paper, we lay the foundation for a novel generation of ad blockers by investigating whether LLM-based or sentence transformer-based approaches can detect generated native ads. To this end, we create the Webis Generated Native Ads~2024 dataset in three steps:
\Ni
we collect 500~keyword queries for each of~10 frequently queried product categories,
\Nii
we collect the corresponding results of the commercial conversational search engines YouChat and Microsoft Copilot, and
\Niii
we use GPT-4 to automatically generate variants of the search results with inserted ads (highlighting a product or brand along with pre-defined qualities).
In our experiments on the new dataset,%
\footnote{Code: \href{https://github.com/webis-de/WWW-24}{github.com/webis-de/WWW-24} \ or \ \href{https://doi.org/10.5281/zenodo.10808158}{doi.org/10.5281/zenodo.10808158}\\
Data: \href{https://doi.org/10.57967/hf/1847}{doi.org/10.57967/hf/1847} \ or \ \href{https://doi.org/10.5281/zenodo.10802427}{doi.org/10.5281/zenodo.10802427}}
the sentence transformers are highly effective at detecting the inserted ads. The LLMs have more difficulties with the task.

\section{Related Work}

Research in the field of search engine advertising mostly focuses on optimizing ads rather than recognizing them. Examples include the automated generation of ad text that is relevant to a query~\cite{duan:2021} and that has a high linguistic quality~\cite{kamigaito:2021}, or the selection of keywords with high expected click-through rates~\cite{bulut:2023,hughes:2019}.

In marketing research, a number of studies focus particularly on the two forms of covert advertising most closely related to our research:
\Ni
native advertising, which imitates the form and appearance of editorial texts~\cite{schauster:2016, wojdynski:2016}, and 
\Nii  
product placement, where paid content relating to products or brands is inserted into media such as films or music videos~\cite{campbell:2019, eyada:2020}.
In both cases, repeated contact with the advertised items is intended to increase familiarity with and preference for them~\cite{storm:2015, avramova:2017}. Interestingly, the effect persists both when people are made aware of the product placement beforehand and with target groups that have a negative attitude towards this form of advertising~\cite{storm:2015}. Particularly relevant for our work is that the effectiveness of ads in textual environments increases with their connectedness to the text content~\cite{olsen:2012, barnhardt:2016}---a ``requirement'' that can probably be outsourced to today's~LLMs.

The persuasive power of~LLMs has been illustrated in several studies. For instance, \citet{spatharioti:2023} showed that people using an LLM-based conversational search engine had a high confidence in the provided information, even when it was incorrect. However, highlighting the potentially false or misleading information in color significantly helped people to recognize them~\cite{spatharioti:2023}. Similar to spotting incorrect information, people also have difficulties spotting native ads in generated search results~\cite{zelch:2024}. Whether some color-based highlighting would be helpful as a first version of ``ad blocking'' or whether the ad effect would still persist was not investigated, though. If highlighted ads would still work, an ad blocker might have to rewrite generated search results to omit a native ad. Still, either highlighting or rewriting require the automatic detection of native ads in the first place.

To remove unwanted advertisements from websites, many web users turn to ad blockers~\cite{shiller:2018}, whose popularity has caused the advertising industry to perceive them as a growing threat~\cite{pujol:2015}. The most common ad-blockers like AdBlock or AdBlock Plus mainly block video ads, pop-up ads, and other forms of online ads~\cite{post:2015}. Some of them work by preventing to load JavaScript files which send requests to ad servers, while others allow loading these scripts but block outgoing requests. Given that these approaches would not detect and block ads woven directly into generated text responses, new solutions are required for generated native ads.

\section{Creating the Webis Generated Native Ads 2024 Dataset}
\label{part3}

To create a realistic dataset of generated native ads, we envision a hypothetical scenario of a provider of a conversational search engine that offers to insert ads into the generated search responses via some instruction-tuned~LLM. Paying advertisers may define
\Ni
queries for which they want an ad to appear,
\Nii
products or brands they wish to advertise, and
\Niii
which qualities of the products or brands should be advertised.
This way, the advertisers are relieved of prompt engineering and the provider is protected from adversaries who might want to inject potentially harmful prompts. Product or brand names and their qualities (adjectives) can be cost-effectively vetted by maintaining controlled vocabularies.

Based on this envisioned scenario, we create the Webis Generated Native Ads~2024 dataset of queries and generated responses with and without inserted native~ads by applying the following steps.

\paragraph{Query selection}
We used ten ``meta topics'' that relate to commercial fields with a range of different products: banking, car, gaming, healthcare, real estate, restaurant, shopping, streaming, vacation, and workout. For each of the meta topics, we collected the 500~most competitive (or all, if fewer are available) keyword queries from the SEO~service \href{https://keyword-tools.org}{keyword-tools.org}, 4,868~keyword queries in total. 

\paragraph{Search response scraping}
We submitted each of the 4,868~queries twice to YouChat%
\footnote{\url{https://you.com}}
and twice to Microsoft Copilot.%
\footnote{\url{https://www.bing.com/search?q=Bing+AI&showconv=1}}
Keeping only English responses with four to twelve sentences, a total of 11,303~responses remained (Table~\ref{table-results}a shows the distribution).

\paragraph{Ad candidate selection.}
To collect simulated requests to the hypothetical native advertising service, for each of the ten meta topics, we asked GPT-4%
\footnote{All mentions of GPT-4 refer to GPT-4 Turbo with knowledge cutoff in April~2023.}
to create a list of suitable products or brands together with short descriptions of to-be-emphasized qualities. We then manually verified, filtered, and expanded the suggestions to create a list of 100~suitable products or brands with to-be-emphasized qualities for each of the ten meta topics.

\paragraph{Ad insertion.}
The actual insertion of the native ads was split into two parts. First, for each query, we asked GPT-4 to select between two and five fitting potential ad candidates from the respective meta topic's product and brand list. From these candidates, we manually selected two suitable items per query, while trying to maximize the overall number of different ads per meta topic. Second, given a query with its original search response, and one of the selected ad candidates with associated qualities, we prompted GPT-4 to integrate the ad candidate and the qualities. GPT-4 was allowed to adapt the qualities lexically while retaining their semantics to increase query fit and vocabulary diversity. From the resulting GPT-4 outputs, we only kept those in which the ad insertion affected a single sentence as GPT-4 sometimes altered other sentences without introducing an ad. The 6,041~remaining search responses with native ads (topic distribution in Table~\ref{table-results}a) were then also added to the Webis Generated Native Ads 2024 dataset. For each native ad, the dataset contains the exact character-wise range of the ad insertion and the character-wise boundaries of the affected sentence.

\paragraph{Validating the inserted ads.}
As an indicator for the lexical similarity of ads, we compute the average \mbox{ROUGE-1} F1-score of pairs of sentences with inserted ads (stop words and advertised item removed, lemmatized). For ads from the the same meta topic, the average~F1 is~7.61, for ads from different meta topics, it is~2.47. This suggests some shared vocabulary but overall the ads seem to be rather lexically diverse.

\begin{table*}[t]%
\small%
\centering%
\caption{%
(a)~Responses per meta topic and search engine. For each search engine, the left (right) column indicates the number of responses without (with) advertisements. The bottom row shows the sum. 
(b)~Detection effectiveness. The results are given for each meta topic in a hold-out test approach. The last row shows the scores for the mixed test set.
(c)~Confidence intervals~(95~\%) for precision and recall across the 11~test sets.
(d)~Illustration of false positives. The highlighted passages are classified as advertising in a response without an inserted ad.}%
\vskip-3ex%
\label{table-results}%
\renewcommand{\tabcolsep}{1.5pt}%
\begin{tabular}[t]{@{}l@{}rrrr@{}}
(a) \\
\toprule
\textbf{Meta} & \multicolumn{2}{c}{\textbf{YouChat}} & \multicolumn{2}{c}{\textbf{Copilot}} \\
\cmidrule(l@{\tabcolsep}r@{\tabcolsep}){2-3}
\cmidrule(l@{\tabcolsep}r@{\tabcolsep}){4-5}
\raisebox{1.25ex}[0em][0em]{\textbf{Topic}} & \footnotesize Orig.\kern-0.3em &\footnotesize Ad &\footnotesize Orig.\kern-0.3em &\footnotesize Ad \\
\midrule
  Banking     & 526 & 248 & 649 & 313 \\
  Car         & 555 & 269 & 851 & 389 \\
  Gaming      & 554 & 323 & 871 & 462 \\
  Healthcare  & 357 & 173 & 655 & 291 \\
  Real estate & 396 & 247 & 599 & 351 \\
  Restaurant  & 467 & 231 & 630 & 331 \\
  Shopping    & 503 & 285 & 791 & 414 \\
  Streaming   & 552 & 296 & 747 & 404 \\
  Vacation    & 359 & 237 & 686 & 398 \\
  Workout     & 148 & 92  & 407 & 287 \\
\midrule
  $\sum|$\,\footnotesize Mixed & 4,417 & 2,401 & 6,886 & 3,640 \\
\bottomrule
\end{tabular}%
\hfill%
\begin{tabular}[t]{@{}ccccc@{\hspace{1em}}ccccc@{}}
\multicolumn{1}{@{}l@{}}{(b)} \\
\toprule
\multicolumn{5}{@{}c@{}}{\textbf{Precision}}        & \multicolumn{5}{@{}c@{}}{\textbf{Recall}}                \\
\cmidrule(r@{1em}){1-5}\cmidrule{6-10}
\footnotesize Alpaca & \footnotesize GPT-4 & \footnotesize Mistral & \kern-0.15em | \footnotesize MiniLM & \footnotesize MPNet & \footnotesize Alpaca & \footnotesize GPT-4 & \footnotesize Mistral & \kern-0.15em | \footnotesize MiniLM & \footnotesize MPNet \\
\midrule
                   0.37 & 0.51 & 0.42 & 0.91          & \textbf{0.95} & 0.43 & 0.82 & 0.43  & 0.89          & \textbf{0.93} \\
                   0.33 & 0.54 & 0.38 & 0.83          & \textbf{0.91} & 0.66 & 0.43 & 0.47  & \textbf{0.99} & 0.99          \\
                   0.35 & 0.48 & 0.42 & 0.86          & \textbf{0.96} & 0.59 & 0.44 & 0.28  & \textbf{0.98} & 0.98          \\
                   0.36 & 0.48 & 0.41 & 0.76          & \textbf{0.88} & 0.41 & 0.85 & 0.37  & 0.99          & \textbf{0.99} \\
                   0.38 & 0.53 & 0.44 & 0.92          & \textbf{0.96} & 0.50 & 0.79 & 0.34  & \textbf{0.99} & \textbf{0.99} \\
                   0.35 & 0.63 & 0.43 & 0.96          & \textbf{0.98} & 0.66 & 0.67 & 0.40  & \textbf{0.96} & 0.95          \\
                   0.37 & 0.53 & 0.42 & 0.89          & \textbf{0.94} & 0.65 & 0.88 & 0.58  & \textbf{0.99} & 0.98          \\
                   0.38 & 0.50 & 0.46 & 0.94          & \textbf{0.97} & 0.60 & 0.73 & 0.48  & 0.92          & \textbf{0.93} \\
                   0.38 & 0.44 & 0.40 & 0.73          & \textbf{0.84} & 0.55 & 0.94 & 0.66  & \textbf{1.00} & \textbf{1.00} \\
                   0.49 & 0.69 & 0.57 & 0.92          & \textbf{0.97} & 0.45 & 0.87 & 0.51  & 0.94          & \textbf{0.98} \\\midrule
\kern-1.25em$|$ \, 0.36 & 0.48 & 0.42 & \textbf{0.99} &         0.98  & 0.54 & 0.77 & 0.49  & 0.91          & \textbf{0.97} \\
\bottomrule
\end{tabular}%
\hfill%
\raisebox{0.2ex}[0em][0em]{(c)\hspace*{-1em}}%
\raisebox{-45.6ex}[0em][0em]{\includegraphics{plot-llm-effectiveness}}%

\medskip
\renewcommand{\tabcolsep}{4pt}
\begin{tabular}[t]{@{}llp{0.799\textwidth}@{}}
(d)\\
\toprule
\bfseries Model     & \bfseries Query   & \bfseries Response \\
\midrule

MPNet                & jetbluevacations      &  JetBlue Vacations is a travel service that offers vacation packages and deals, including flights, hotels, car rentals, and activities in hundreds of destinations around the world. [\ldots] \hlc[yellow!35]{JetBlue Vacations makes it easier for travelers to book their flights and hotels at the same time, providing a seamless planning process for a convenient travel experience.} [\ldots].
\\
MiniLM                & ladies shorts      & When it comes to women's shorts, there are various styles and materials to choose from [\ldots] \hlc[yellow!35]{Overall, women's shorts cater to a wide range of preferences, from casual and laid-back looks to more stylish and elegant options, ensuring there's something for everyone.}
\\
GPT-4                & synchrony home      & \hlc[yellow!35]{Synchrony Home is a credit card offered by Synchrony Bank that is specifically designed for making home-related purchases.} [\ldots] The Synchrony Home Credit Card provides promotional financing options, [\ldots]
\hlc[yellow!35]{Synchrony Bank offers a range of financial services, including savings accounts, CDs, money market accounts, IRAs,} [\ldots].
\\
GPT-4            & t shirts for women & 
Here are some popular women's t-shirts that you might like:
\hlc[yellow!35]{Levi's Perfect T-Shirt: This white t-shirt is made of 100\% cotton and} [\ldots] \hlc[yellow!35]{ASOS Women's T-Shirts \& Vests: ASOS has a wide range of women's t-shirts and vests} [\ldots]
\\
\bottomrule
\end{tabular}
\end{table*}

\paragraph{Dataset split.}
The Webis Generated Native Ads 2024 dataset comes with a fixed split of a 70\%~subset for training, 15\%~for validation, and 15\%~for testing. To avoid information leakage between the subsets, all instances with the same advertised item are in the same subset, while simultaneously minimizing the overlap of queries between the different subsets. In addition to the fixed split, ten straightforward hold-out versions can be created similar to a cross-validation setup by treating each meta topic once as the test set, and the nine remaining topics as training and validation sets.

\section{Detecting Generated Native Ads}
\label{part4}

We have developed basic approaches to detect native ads. The first approaches are based on pre-trained sentence transformers, namely \emph{all-MiniLM-L6-v2} (MiniLM)%
\footnote{\url{https://huggingface.co/sentence-transformers/all-MiniLM-L6-v2}}
and \emph{all-mpnet-base-v2} (MPNet)%
\footnote{\url{https://huggingface.co/sentence-transformers/all-mpnet-base-v2}}
with an additional linear layer on top of the embeddings. We fine-tune these models for next sentence prediction on sentence pairs. A positive pair consists of a sentence with an inserted ad and one of its immediate neighbors. For each positive pair, the corresponding negative pair uses the original sentence without ad. To match the label distribution of our dataset (fewer responses with ads), we sample additional negative pairs from the remaining sentence pairs of the original responses. Using the Adam optimizer with binary cross-entropy loss with batch sizes of~48 and~16 and learning rates of~1$e$-5 and~5$e$-6 for \emph{MiniLM} and \emph{MPNet}, respectively, we fine-tune eleven versions of both models, one per hold-out meta topic, and one on the mixed split containing responses from all ten meta topics. As final weights, we select the best validation loss over 30~epochs.

The other approaches are based on three instruction-tuned LLMs in a zero-shot setting: GPT-4, Mistral-7B-Instruct, and our own reproduced version of Alpaca~7B. We prompt GPT-4 and Mistral with a query and its respective response with and without inserted ad, asking each to return products/brands and passages identified as advertising. Since Alpaca has difficulties to respond to the full prompt, we only ask it to return products/brands.

\section{Evaluation}
\label{part5}

The effectiveness scores of the sentence transformers (\emph{MiniLM} and \emph{MPNet}) and of the~LLMs on each of the eleven test sets are given in Table~\ref{table-results}b and~c. The sentence transformers generally achieve precision and recall values above~0.8 (MiniLM) and above~0.9 (MPNet). The only exceptions are the meta topics healthcare and vacation with precision scores of~0.76 and~0.73 for \emph{MiniLM} and~0.88 and~0.84 for \emph{MPNet}. Their false negatives are almost exclusively responses in which the inserted ad has a close relation to the query, such as advertising ``PNC Virtual Wallet'' for the query {\tt\small pnc\,online}. The false positives are more diverse but tend to focus on a specific kind of vocabulary as illustrated by two examples in Table~\ref{table-results}d: while the response to {\tt\small jetbluevacations} features advertisement-like language about the brand by that name, the response to {\tt\small ladies shorts} has a similar style without explicitly mentioning a product or brand.

Compared to the sentence transformers, the tested~LLMs achieve a lot lower precision and recall values. Among the~LLMs, GPT-4 is the most effective and Alpaca often has a higher recall than Mistral, while the latter has a higher precision. GPT-4 and Alpaca tend to achieve higher recall than precision. A majority vote of the three~LLMs would increase the recall on the mixed test set to~0.90 at a precision of~0.41. Analyzing the false positive examples reveals that they stem from the queries having a commercial character (see Table~\ref{table-results}d). For the query {\tt\small synchrony\,home}, GPT-4 classifies both the explanation of the credit card as well as the list of Synchrony Bank's offerings as ads. While the former directly relates to the query, the latter can be argued to go beyond that and have advertising character. The query {\tt\small t\,shirts\,for\,women} illustrates another pattern in which the LLMs classify the returned list of products as ads. Again, it is a question of personal judgment if lists of products in response to commercial queries are considered as ads or not.

To more systematically analyze the predictions of GPT-4 (most effective~LLM), we sample 50~false positive and 50~false negative examples and let three annotators assign manual labels to them without revealing which is which. The few disagreements have been resolved by majority vote, which indicates that the perception of advertising language is at least somewhat subjective. Only three of GPT-4's false negatives agree are also false negatives of the human annotators. In contrast, for the model's false positive predictions, the majority of the annotators also agreed with GPT-4 in~26 of the 50~cases (seven with perfect inter-annotator agreement). This shows that some of the original responses from YouChat and Copilot already use advertising language prior to any insertions. 

\section{Discussion and Limitations}
\label{part6}

Besides the intended positive, colorful language of an inserted ad, the sentence transformers also pick up another pattern: If GPT-4 finds no ``natural'' relation between an ad and a response text, it uses expressions like ``alternatively'' (54~cases) or ``for those who'' (399~cases) as an introduction. Hence, our results are limited to \mbox{GPT-4's} current ``advertising style'' and our ad insertion prompt. With access to organic pairs of queries and ads, future research can further explore the generalizability of sentence transformers.

A manual analysis of the false positives reveals that advertising language is already present in some responses prior to our insertions. For queries containing product or brand names, the search results can include websites by the corresponding companies, describing the item in a positive, advertising manner. As the results define the context of the conversational search engine, it occasionally reproduces their style in its response. These sentences are often classified as ads by both LLMs and sentence transformers. While these predictions reduce the precision scores, we consider them as correct in the context of ad blocking. Future research should explore the detection of ads that are not introduced externally, but from the search results. Chain-of-thought-prompting might be useful in this context as it provides reasoning for a classification.
\section{Conclusion}

In this paper, we present a first approach to detect native ads in generated responses of conversational search engines. Experiments on our newly created Webis Generated Native Ads~2024 dataset of responses for search queries from YouChat and Microsoft Copilot, where GPT-4 inserted native ads show that sentence transformers can reliably identify the ads. This suggests that LLM-generated native ads currently follow an underlying pattern that ad blocking systems can exploit to highlight or block detected native ads. In a systematic evaluation of false positive detections, we further found that the organic YouChat and Microsoft Copilot responses also already contain some advertising-like language. This happens especially when text from official websites of some brand is reused without further adaptation.

Our study demonstrates the feasibility of generated native ads as well as that of detecting them at client side. Even if commercial conversational search engines would tap into this revenue source, there is potential to defend against it. In future work, we plan on extending our study to cover more diverse types of native ads.

\begin{acks}
This publication has been partially supported by the European Union's Horizon Europe research and innovation programme under grant agreement No~101070014 (OpenWebSearch.EU).
\end{acks}

\begin{raggedright}

\end{raggedright}
\end{document}